%% file: main.tex
\documentclass[sigconf]{acmart}

\usepackage{multirow}
\usepackage{textcomp}
\usepackage{todonotes}
\AtBeginDocument{%
  }

\copyrightyear{2024}
\acmYear{2024}

\setcopyright{acmlicensed}\acmConference[CHASE '24]{2024 IEEE/ACM 17th International Conference on Cooperative and Human Aspects of Software Engineering }{April 14--15, 2024}{Lisbon, Portugal}

\acmBooktitle{2024 IEEE/ACM 17th International Conference on Cooperative and Human Aspects of Software Engineering (CHASE '24), April 14--15, 2024, Lisbon, Portugal}

\acmDOI{10.1145/3641822.3641880}
\acmISBN{979-8-4007-0533-5/24/04}

\begin{document}

\title[Accountability in Software Engineering]{Understanding the Building Blocks of \\Accountability in Software Engineering}

\author{Adam Alami}
\affiliation{%
  \institution{Aalborg University}
  \country{Denmark}
}

\author{Neil Ernst}
\affiliation{%
  \institution{University of Victoria}
  \country{Canada}}

\renewcommand{\shortauthors}{Alami and Ernst}

\begin{abstract}
   
  In the social and organizational sciences, accountability has been linked to the efficient operation of organizations. However, it has received limited attention in software engineering (SE) research, in spite of its central role in the most popular software development methods (e.g., Scrum). In this article, we explore the mechanisms of accountability in SE environments. We investigate the factors that foster software engineers' individual accountability within their teams through an interview study with 12 people. Our findings recognize two primary forms of accountability shaping software engineers individual senses of accountability: \emph{institutionalized} and \emph{grassroots}. While the former is directed by formal processes and mechanisms, like performance reviews, grassroots accountability arises organically within teams, driven by factors such as peers' expectations and intrinsic motivation. This organic form cultivates a shared sense of collective responsibility, emanating from shared team standards and individual engineers' inner commitment to their personal, professional values, and self-set standards. While institutionalized accountability relies on traditional ``carrot and stick'' approaches, such as financial incentives or denial of promotions, grassroots accountability operates on reciprocity with peers and intrinsic motivations, like maintaining one's reputation in the team. 
  
\end{abstract}


\maketitle

\input{introduction}

\input{theory}

\input{related}

\input{methods}

\input{findings}

\input{discussion}

\input{validity}
\input{conclusion}

\bibliographystyle{ACM-Reference-Format}
\bibliography{references}

\end{document}

%% file: introduction.tex
\section{Introduction}\label{sec:introduction}

Throughout history, accountability has been acknowledged as an enabler for the survival, stability, and longevity of social systems \citep{gelfand2004culture}. Greek philosophers, such as Aristotle, Plato, and Zeno, discussed the concept within the context of justice, punishment, and social control \citep{schlenker1994triangle}. In modern literature, accountability has been linked to the efficient functioning of organizations \citep{gelfand2004culture}. In their pioneering work Katz and Kahn outlined the significant effort made by organizations in implementing control mechanisms to reduce the uncertainty of human behavior to establish more predictable pattern of activity \citep{katz1978social}.

\textbf{Accountability is the expectation of being held responsible for one's actions and the need to provide explanations and reasoning for such actions to others in the future} \citep{frink2004advancing}. Organizational history is littered with accountability failures, from the flawed mirror in the Hubble telescope to the infamous collapse of the Arthur Andersen accounting firm in the wake of the Enron and WorldCom scandals \citep{frink2004advancing}. 

Information technology (IT) and software engineering (SE) are not exempt from accountability failures. The Boeing 737 MAX software issue and Healthcare.gov launch have been attributed to some extent to a lapse in accountability \citep{johnston2019boeing,angelo2015healthcare}. Both crashes of the Boeing 737 MAX in Indonesia in 2018 and Ethiopia in 2019 were linked to the MCAS software \citep{johnston2019boeing}. Investigation reports claimed that the software had design and implementation issues \citep{johnston2019boeing}. This case underscored accountability failures within Boeing's software development processes, including inadequate quality assurance practices and a lack of documentation \citep{johnston2019boeing}. Similar conclusion have been drawn in the case of the launch of Healthcare.gov in 2013 \citep{angelo2015healthcare}. The website was marred by not only usability issues but also technical and functionality defects \citep{angelo2015healthcare}.

Social systems rely on expectations to govern human behavior \citep{frink2004advancing}, irrespective of organizational complexity. Conformity to expectations is encouraged through observation, evaluation, and imposing penalties or rewards \citep{frink2004advancing}. Social systems functioning and integrity are partly sustained by the mechanisms of accountability, particularly in the context of formal institutions \citep{frink2004advancing}. However, while mechanisms of accountability carry deep and intricate implications for fundamental organizational operations, it has received less scholarly attention from the SE research community.

Studies have demonstrated that perceptions of accountability contribute significantly to the effectiveness of holding individuals accountable. Mero et al. found that decision makers feeling accountable to management in higher hierarchy of the organization perform within the expected standards and exhibit a greater degree of alignment with organizational goals \citep{mero2007accountability}. Wallace et al. reported that felt accountability among restaurant managers is conducive to an environment of psychological empowerment resulting in enhanced performance, i.e., increased store sales and store service quality \citep{wallace2011structural}. Finally, Mero et al. also found that sales staff perception of accountability for sales target had significant effect on their performance \citep{mero2014field}.

In SE, accountability has been acknowledged in pioneers' work as far back as the 80s. In Boehm's paper on the ``seven basic principles of software engineering,'' he proposed one principle for ensuring accountability for outcomes, i.e., ``maintain clear accountability for results'' \citep{boehm1983seven}. Since then, the concept received little attention until the Agile movement. For example, ``accountability'' appears nine times in the Scrum guide \citep{schwaber2011scrum}. The guide has a strong emphasis on accountability as a control mechanism, for example, ``holding each other [developers and team members] accountable as professionals'' \citep{schwaber2011scrum}. But we still know little about the mechanisms used in SE to exercise accountability, nor do we understand which outcomes organizations seek to control. Therefore, we ask

\medskip

\noindent \textbf{RQ1:} In software engineering environments, what mechanisms do organizations intentionally implement to establish individual accountability, and what outcomes are they aiming to control?

\medskip

\noindent \textbf{RQ2:} What are the key factors and practices that contribute to the development of individual accountability, within software engineering teams, uninfluenced by externally imposed organizational mandates?

\medskip

\textbf{RQ1} explores the tools, practices, and structures purposefully implemented by organizations as mechanisms to ensure individual accountability amongst software engineers. This question enables an investigation into the intentions behind the implementation of these mechanisms and an understanding of the specific outcomes they seek to manage or control. Conversely, \textbf{RQ2} delves into the cultivation of individual accountability among software engineers within a team context, independent and uninfluenced by mechanisms instituted by the organization or inherent in SE processes. 

By posing this question, we aim to probe deeper into the interpersonal and individual aspects within SE teams, exploring personal attributes, team dynamics, and organizational culture that foster individual accountability. This design allowed us to identify and compared two types of accountability: one formally designed by the organization and externally imposed, and the other team-inspired and team-influenced accountability.

In this study, our focus is the micro-level or \emph{felt accountability}, which is the \textbf{individual's perception of accountability} \citep{frink1998toward,hall2017accountability}. Katz and Kahn's ``roles theory'' posits that individuals behave according to the expectations of their defined roles, influencing their perceptions, behaviors, and actions, including their sense of accountability \citep{katz1978social,frink2004advancing}. Roles theory recommends that individuals' accountability should be interpreted based on roles \citep{frink2004advancing} (see Sect. \ref{sec:theory} for further discussion). Accordingly, the scope of our study is also a role-centric investigation into individual accountability. We opted to focus on software engineers because they shape critical SE outcomes like security and quality. This by no means undermines other roles like quality assurance analysts or testers, but rather remains a theoretically grounded decision.

To meet the objectives we set, we carried out a qualitative interview study with 12 interviewees. We opted to explore the topic qualitatively using interviews to gain contextualization to SE, and identify alignments and delineations for the existing theoretical framework. This approach has yield insightful results. While some of our findings are in-line with existing theories, we also show that SE has particularities not captured in previous theories. This approach has not only validated some aspects of the prevailing theories but has also enriched the theoretical landscape with SE-specific insights. 

For example, while some of our findings related to \textbf{RQ1} are well aligned with existing theories, this was not the case for \textbf{RQ2}. The findings of \textbf{RQ2} show that certain factors and practices fostering individual accountability in SE teams emerge naturally and are not always the result of externally imposed mandates (see Sect. \ref{sec:discussion} for further discussion in our theoretical contribution).
Our findings:
\medskip

\noindent \textbf{RQ1:} We found that \textit{institutionalized} accountability often employs traditional approaches to promote and control a sense of individual accountability, such as financial incentives or denial of promotions. Strategies of institutionalized accountability shape engineers' behaviors to align with desired outcomes. 

These strategies range from incentives like financial bonuses to deterrents that underscore potential repercussions for undesired behaviors. These accountability drivers are purposefully designed to channel the influence of different strategies to achieve outcomes like software quality and satisfying deadlines.

\medskip
    
\noindent \textbf{RQ2:} We identified \emph{grassroots} accountability, which is underpinned by software engineers' feeling of reciprocity to their peers, to meet collective expectations, and their intrinsic motivations. Our data show that software engineers internalize their peers' expectations to contribute to a collective sense of responsibility towards team's outcomes. 

The feeling of accountability towards the outcomes is also reflecting both engineers' personal standards and their desire to maintain a favorable professional reputation. The dynamics of grassroots accountability are augmented by software engineering practices, like testing, code review, and peer feedback, and the nurturing of psychological safety within teams. We found that both institutionalized and grassroots accountability are steered toward driving software engineers to meet established expectations for software quality, code quality, software security, and meeting project's deadlines.

Our study deepens the understanding of theoretical perspective of accountability in the context of SE, and provides practical guidelines for organizations seeking to enhance accountability among their engineering teams. We contribute by providing an elaboration of accountability in the context of SE, and we identify a dichotomy of accountability in SE. This dichotomy may imply that peers' expectations and intrinsic drives could be equally effective in promoting accountability as formalized strategies. Our findings also show that certain SE practices, like testing, and code review can nurture and augment both institutionalized and grassroots accountability. Even though, intended as engineering practices, our findings indicate that they are not just for quality assurance but they can also foster a sense of accountability.

%% file: theory.tex
\section{Theoretical Background}\label{sec:theory}

\noindent Accountability is considered the link between individuals and their social system, creating an identity relationship associating individuals with their actions and performance \citep{tetlock1992impact,mero2007accountability}. Accountability has implications at all organizational levels \citep{frink1998toward}. However, our focus in this study is the micro-level or \emph{felt accountability} (referred to simply as ''accountability''), which refers to the individual's perception of accountability \citep{frink1998toward}; the most studied type of accountability \citep{hall2017accountability}. This type of accountability is contingent upon the individual's own interpretation \citep{frink1998toward}, rather than the external milieu-imposed interpretations \citep{folger2001fairness}.

Compliance with standards is evaluated at various layers of social systems, including the individual, the dyad, the group, the organization, and society as a whole \citep{gelfand2004culture,frink1998toward}. As such, we adopted Gelfand et al.'s definition of accountability, which \emph{``the perception of being answerable for actions or decisions, in accordance with interpersonal, social, and structural contingencies, all of which are embedded in particular sociocultural contexts''} \citep{gelfand2004culture}. This definition emphasizes the perceptual nature of accountability acknowledging that standards and their violations are socially constructed and informed by sensemaking processes \citep{gelfand2004culture,frink1998toward}. The concept of accountability inherently implies an anticipated evaluation \citep{hall2017accountability}. For the latter to take place, the individual must engage in an account-giving process \citep{frink2008meso}, which may result in rewards or sanctions.

Frink and Klimoski suggest that any conceptualization of workplace accountability should consider both the formal and informal manifestations of accountability \citep{frink1998toward}. While formal accountability is formalized over time and grounded in organizational rules and policies \citep{frink2008meso}, informal is the perceived accountability ``outside of or beyond formal position or organizational policy'' \citep{zellars2011accountability}. Informal accountability is grounded in unofficial expectations and discretionary behaviors that result from the socialization of network members \citep{romzek2012preliminary}. Shared norms also lay out an informal code of conduct used by group members as a reference for appropriate and inappropriate behaviors \citep{romzek2012preliminary,romzek2014informal}.

Gelfand et al. contend that once individuals assimilate a culture's practices and values through the process of enculturation, they develop ``cognitive maps'' delineating their accountabilities to other individuals, groups, and organizations in their sociocultural context \citep{gelfand2004culture}. Such cognitive maps lay out the specifications of how individuals are answerable within the accountability web \citep{gelfand2004culture,frink1998toward}.

Roles theory posits that work units develop norms that establish appropriate behaviors and actions based on the division of labor through the concept of roles \citep{katz1978social}. Roles inherently prescribe behaviors and expectations, aiming to create structure. Role expectations are standards against which individual actions and behaviors are evaluated \citep{katz1978social}. Frink and Klimoski suggest that a role-centric approach to accountability recognizes the role-specific expectations set for individuals \citep{frink2004advancing}. Roles are lenses through which expectations are set and managed. These expectations are used to evaluate individuals' performance in account-giving processes \citep{frink2004advancing}. In this view, roles are not merely specifications of duties but rather complex frameworks within which accountability is understood, controlled, and evaluated.

%% file: related.tex
\section{Related Work}\label{sec:related}

Our search for literature on the topic of accountability specific to the context of SE shows that there remains a gap in our understanding of how accountability influences SE outcomes. There is substantial research focused on the accountability of software and technology to its users \citep{hutchinson2021towards,nissenbaum1996accountability}, e.g., in keeping users safe from accidents like the Therac-25 disaster \citep{mcquaid2012software,jugovic2023case}. However, there is a scarcity of research exploring accountability from the perspective of software engineers, particularly regarding how they perceive and internalize accountability, the reasons behind this sense of responsibility, and the specific objectives towards which this accountability is directed.

In behavioral SE, several studies ~\citep{Graziotin2017OnUnhappinessSoftware,Graziotin2018WhatHappensWhen} have looked at the connection between productivity and developer or team happiness. Graziotin ~\cite{Graziotin2017OnUnhappinessSoftware} for example, has connected developer happiness to productivity, and some of the causal factors of happiness (or unhappiness) include mechanisms related to accountability, e.g., under-performing colleagues. However, these studies do not draw connections to accountability explicitly, and focus mostly on individual emotions rather than team dynamics.

We found two further research studies, one on agile team perception of accountability \citep{burga2022examining}, and the second on the influence of Scrum practices on quality \citep{alami2022scrum}.

Burga et al. studied nine software development teams as they adopted agile practices to understand how they influence the team's members accountability \citep{burga2022examining}. The study found that in agile teams, expectations are not always institutionalized, and team members develop their understanding of their roles and corresponding accountabilities throughout the adoption process \citep{burga2022examining}. The researchers further explain that agile practices such as stand-ups, sprints, and story points enable human behavior conducive to feeling accountable \citep{burga2022examining}. The findings also suggest that accountability within agile teams is a shared concept, with different team members taking ownership of various tasks and the team collectively accountable for outcomes \citep{burga2022examining}. Our work shows that, irrespective of whether the team is adhering to agile practices or not, the sense of collective accountability may grow organically when teams share their expectations.

Alami and Krancher studied the extent to which Scrum practices influence SE teams to achieve software quality \citep{alami2022scrum}. The results suggest that software developers feel individually accountable for the quality of their deliverables in Scrum environments. The study's participants explained that self-governing and the empowerment experienced in working in Scrum compel them to take ownership of their work and feel accountable for its quality. This sense of accountability is shared and collectively adhered to within the team. The consequences of failing to be accountable for one's work include a loss of peer recognition \citep{alami2022scrum}. While the study demonstrates that the Scrum environment fosters a sense of accountability, it does not delve into the specific accountability mechanisms used to control it or how they shape behaviors and outcomes.

Our study focuses on a broader perspective of accountability in SE, encompassing various drivers, influencers, and different mechanisms of control and how collectively they contribute to achieving multiple outcomes. In contrast, the literature seems primarily interested in the perception and influence of accountability within agile teams without exploring the more granular mechanisms of control or a wide range of outcomes.

%% file: methods.tex
\section{Methods}\label{sec:methods}

\begin{table*}[t!]

  \begin{center}
    \footnotesize
    \caption{Interviewees' charachterestics.}
    \label{tbl:population}
    \renewcommand\arraystretch{0.80}
        
    \begin{tabular}{l|p{2.5cm}|c|c|p{3cm}|p{3.5cm}|c|p{1.1cm}}
      \hline
      \textbf{\#} & \textbf{Role} & \textbf{Exp.} & \textbf{Method} & \textbf{Project type} & \textbf{Industry sector} & \textbf{Gender} & \textbf{Country}\\
      \hline
    
        P1 & Software Engineer & 3-5 years & DevOps & Media data platform & Information Technology services & Male & Germany\\
        P2 & Software Engineer & 3-5 years & Hybrid & Utility software & Information Technology services & Male & UK\\
        P3 & Software developer & 9-11 years & Scrum & Robotics software & Robotics manufacturing & Male & USA\\
        P4 & Software developer & 9-11 years & Hybrid & Business intelligence & Information Technology services & Male & Italy\\
        P5 & Sr. software engineer & 6-8 years & Scrum & Online market place & Information Technology services & Non-binary / third gender & Germany\\
        P6 & Sr. software engineer & $>$12 years & Kanban & Infrastructure migration & Banking services & Male & Canada\\
        P7 & Software developer & $<$3 years & Hybrid & Embedded software & Information Technology services & Male & France\\
        P8 & Sr. software engineer & $>$12 years & Scrum & Data migration & Information Technology services & Female & India\\
        P9 & Sr. software engineer & 6-8 years & Scrum & CRM software & Information Technology services & Male & Serbia\\
        P10 & Sr. software engineer & 9-11 years & Hybrid & Data management & Global software vendor & Male & Canada\\
        P11 & Software developer & 3-5 years & Scrum & Telecommunication software & Global software vendor & Male & UK\\
        P12 & Software engineer & $<$3 years & DevOps & Process automation & Information Technology services & Female & India\\

     \bottomrule
     
    \end{tabular}
   
  \end{center}
  
\end{table*}

We carried out a qualitative interview study with 12 interviewees. 
We opted to use interviews because they are particularly well suited to uncovering the complex social nuances and personal experiences of individuals feelings of accountability. They are also adept at providing a more contextual understanding of accountability within the specific context of SE. Interviews allowed us to access the emotional perspective of our interviewees and the insights that stem from their direct experiences. By tapping into these experiences, we aim to identify potential areas where existing theories can be refined and extended to better reflect the context of SE.

Our research design is also anchored in the established theoretical framework proposed by Frink and Klimoski \cite{frink1998toward}. However, the framework may not fully encapsulate the unique aspects of accountability particular to SE environments. Therefore, we aimed to combine theoretical perspectives with the empirical data gathered through interviews. This integration of theory and the empirical knowledge of practitioners has yielded findings contributing to both the theoretical discourse and the contextual understanding of accountability in SE. We highlighted our findings with theoretical propositions to identify alignments and discrepancies. Upon the completion of our inductive analysis, we mapped our findings to the existing theoretical framework (Sect. \ref{sec:theory}) to identify support for similarities and discrepancies in our findings. This exercise allowed us to identify where the theory resonates with the experiences of software engineers and where it diverges (see Sect. \ref{sec:discussion} for further discussion).

\subsection{Interviewee selection} We used Prolific\footnote{\url{https://www.prolific.co/}}, a research market platform, to recruit participants for the interviews. We used a list of 365 software engineers, Tech leads, software architects, etc. curated from a previous study \cite{alami2023antecedents}, where participants were vetted to ensure the appropriateness of their background as software professionals. We used the list because of the robust pre-screening process we used, which has yielded qualified participants.

In our previous study \cite{alami2023antecedents}, our participant list consisted exclusively of individuals working in agile software development environments. To ensure a diverse sample that includes non-agile perspectives, we conducted a second pre-screening exercise. We collected 200 entries, of which 197 qualified. We used a free-text question and asked our participants to program a small function and describe how they overcame a challenge in their current role. We scrutinized the answers, and we used ChatGPT for content AI detection. Two answers did not pass the detection, and one answer was an incomprehensible text.

From our previous study list \cite{alami2023antecedents} and the new pre-screening, we curated a total of 562 qualified participants. Then, we sent an invitation to the 562 participants to participate in the pre-screening survey\footnote{Both pre-screening surveys for non-agile participants and interviewees are available in the replication package} for the interview selection. The purpose of this additional pre-screening is to collect data on how accountability mechanisms shape the participant's work environment, which we did not have in previous pre-screening data.

Based on the 159 responses we received, we chose twenty participants; however only twelve accepted to participate. We used a comprehensive set of criteria to select our interviewees, spanning factors such as country of residence, age, experience, role, gender, education, software development method (e.g., agile, plan-driven, etc.), team size, project scope (i.e., what the SE team is developing), type of development (e.g., custom or maintenance), in-house development or outsourcing environments, and accountability practice within the participant team and organization.

We wanted participants with varying degrees of accountability and practices used to control it, reinforcing the depth and breadth of our sample. For example, among our interviewees (P1, P2, and P3, as listed in Tbl. \ref{tbl:population}), all identified ``code quality'' as their primary individual accountability within their respective teams. However, they reported varying levels of accountability. P1 expressed a high level of accountability, responding with ``strongly agree'' to all accountability-related questions. In contrast, P2 and P3 displayed a moderate level of accountability, mostly responding with ``somewhat agree'' to most of the accountability questions.

\begin{table*}[t!]
\footnotesize
    \caption{Example of \textbf{RQ1} \& \textbf{RQ2} Pattern Codes}

    \label{tab:themes}
    \renewcommand\arraystretch{1.8}
    
    \begin{tabular}{p{4.1cm}p{2.8cm}p{9.5cm}}
    \toprule
      \textbf{Pattern codes} & \textbf{First Cycle codes} & \textbf{Examples from the data}\\
      \hline

      \multicolumn{3}{l}{\textbf{RQ1}} \\\hline
      
      \multirow{5}{*}{\textbf{Outcomes}} & Code quality  & \emph{``So I have to write good quality code, good code, the good code, which is working and my teammates are happy with''} (P11).\\
      & Software security & \emph{``If something, for example, obscures that our secure security critical security critical, then I think I would feel guilty''} (P1). \\
      & Meeting deadlines  & \emph{``... the times that you have to develop something if you are not on the timeline ... but you have written the code good quality ... then it is not okay.''} (P11). \\
      \hline
      
      \multirow{3}{4.1cm}{\textbf{Institutionalized accountability drivers}} & Financial rewards  & \emph{``So there's extra money, to be honest, is just encouraging me to do more. And also encouraging me to do the job by right'' P(2).} \\
      & Punishment  & \emph{``I think there is requirements of bit of both punishment without punishment, team cannot survive ... you have to set the rules on this is the threshold of punishment ... reward is always required. ''} P(10).  \\ \hline

      \multicolumn{3}{l}{\textbf{RQ2}} \\\hline

      \multirow{3}{3cm}{\textbf{Peers' expectations}} & Meeting peers' expectations & \emph{``Normally I'd saw my normal behaviors, to, to try to be compliant ... I would still do it for the [code review] approval sake'' (P5).} \\
      & Motivational pressure & \emph{``These ongoing reviews pressure me to meet their [his peers'] expectations.''} (P2). \\
      \hline

      \multirow{3}{*}{\textbf{Intrinsic drivers}} &  Personal integrity & \emph{``Yeah. Your integrity will give you this professional image that will you preserve for the rest of your life, I guess''} (P7). \\
      & Feeling shame & \emph{``It's a basic stuff that needs to be done. But not doing think right, but would be shameful.'' (P7).}  \\
      
     \bottomrule
     
    \end{tabular}
  
\end{table*}

The pre-screening for interviewees took place in the first week of September 2023. Table \ref{tbl:population} documents our interviewees. ``Exp.'' is the participant's work experience in software development. ``Methods'' are the software development methods used in the participant's team, and ``hybrid'' refers to combining a plan-driven and an agile method. ``Project type'' describes the software product being developed by the participant's team. ``Industry sector'' specifies whether the interviewee's employer operates in technology, finance, healthcare, manufacturing, or other sectors. We paid \textsterling30 to each interviewee.

Notably, our sample is geared towards software engineers. This choice is methodologically driven and theoretically grounded. Limiting our sample to a specific role reduces the potential for variations that may be introduced by the inclusion of multiple roles. This narrow focus strengthens the internal validity of our study and allows for role-centric conclusions. Roles theory suggests that individuals' accountability is closely linked to their roles and responsibilities within an organization \citep{katz1978social,frink2004advancing} and individual accountability should be interpreted based on roles \citep{frink2004advancing}.

\subsection{Data collection} We used semi-structured interviews. Although we used a predefined interview guide (see our replication package for the complete guide), semi-structured interviews are flexible and allow aligning the questions with the flow of the interview by prompting the participants to disclose further relevant details based on the topic being discussed.

We structured the guide into five sections: after the introduction to the interview, section one sought to capture data on the background of the interviewee and a general understanding of how accountability manifests in their teams. Section Two sets out to identify accountability mechanisms and the outcomes they seek to control in the interviewee's team. Section Three delves into how the discussed accountability mechanisms impact the desired outcomes. Section Four aimed to reveal any moderating factors or conditions (e.g., roles or interpersonal relations at work) that may influence the effectiveness of accountability mechanisms. Before concluding the interview, Section Five identified best practices used by the interviewee's team to implement effective accountability mechanisms.

Our interviewees were distributed geographically, so all interviews were conducted using Zoom and lasted 40--60 minutes with a total of 10h32min of audio. The audio generated a total of 152 pages verbatim after transcribing (an average of 12 pages per interview). The first author conducted the interviews in September–October 2023. We used Otter.ai\footnote{\url{https://otter.ai/}} an online transcription tool to transcribe the audio recordings.

\subsection{Data analysis}
Our analysis process was iterative; we analyzed the data as we progressed with the interviews to monitor saturation. We used guidelines from Miles, et al.~\citep{miles2014qualitative} and Salda{\~n}a~\citep{saldana2021coding} to analyze the interview data. The guidelines encompass two phases: (1) \textit{First Cycle} and (2) \textit{Second Cycle} \citep{miles2014qualitative,saldana2021coding}. They also recommend several techniques that facilitate finding relationships among themes \citep{saldana2021coding}.

\paragraph{First Cycle.} During this phase of coding, data ``chunks'' are assigned meaningful labels or codes using several coding types like ``descriptive coding'', ``In Vivo'', ``process coding'', and ``causation coding'' \citep{miles2014qualitative}. The coding types help researchers to interpret and classify the data, and they capture different dimensions of the qualitative data \citep{miles2014qualitative}. For example, ``descriptive coding'' describes the data authentically; it is a straightforward labeling of the data based on its surface-level attributes \citep{miles2014qualitative}. ``Process coding,'' on the other hand, allows capturing processes expressed in the form of sequences and actions in the data \citep{miles2014qualitative}. We used an inductive coding approach. This decision allowed us to gain data-driven insights. The inductive approach allowed us to derive codes directly from the data without imposing any preconceived notions. This is relevant in the context of our study, given that our objectives are to contextualize existing theories to SE, identify particularities unique to SE, and provide eventual delineations for the theoretical framework of Frink and Klimoski \cite{frink1998toward}.

The first author conducted the first phase of coding. Subsequently, the second author reviewed the codes and provided feedback and suggestions for additional codes and re-labelling of some of the codes. Then, the first author reviewed and decided on a final list of codes; this allowed us to tune the first coding round and consolidate our decisions on codes. This ``reliability check'' was an opportunity to reconcile our differences for more credible analytical conclusions \citep{miles2014qualitative}. 

\paragraph{Second Cycle.} In this second phase of coding, the codes of the First Cycle are synthesized into a concise, unified scheme, encompassing several themes and their relationships \citep{saldana2021coding}. A ``pattern coding'' exercise takes place to cluster the codes into Pattern Codes based on either their code types, pertaining to similar patterns or meaning in the data, or logically sharing some characteristics \citep{miles2014qualitative}. This systematic approach allows for transforming a micro-level examination of codes into a macro-level understanding of the data. Similarly to the previous phase, the first author carried out this activity. Then, the second author reviewed and provided feedback to reach consensus. Table \ref{tab:themes} documents some examples of our themes and their corresponding codes related to our research questions. We commenced this cycle right after each interview's codes from the First Cycle became available, which allowed us to monitor for saturation \citep{morse2004theoretical,aldiabat2018data}.

\paragraph{Pattern Codes relationships.} Pattern Codes relationships (see Fig. \ref{fig:conceptual}) are the connections or links that we discovered between our Pattern Codes \citep{saldana2021coding}. After we finalized our Pattern Codes, we sought to uncover the relationships among them to underline patterns and structures in our data. The purpose of this activity is to understand the relationships between them and how they may influence or relate to one another. We used a ``causal network'' analysis \citep{miles2014qualitative} to coordinate connections between the ``Pattern codes'' and how they relate to one another \citep{miles2014qualitative}.

We monitored data \emph{saturation} \citep{morse2004theoretical,aldiabat2018data} throughout our analysis process. We began the analysis as soon as the interview transcript became available. This process allowed us to ensure the adequacy of our sample size \citep{aldiabat2018data}. By continuously comparing data and emerging themes and switching back and forth between data collection and analysis \citep{bowen2008naturalistic}. Some of our Pattern Codes reoccur strongly in the data, e.g., intrinsic motives and peers' expectations (12 times), and others only five times, e.g., psychological safety. We documented this process in our replication package.

\paragraph*{Replication package.} We share our data and analysis artifacts at this \href{https://doi.org/10.5281/zenodo.10105278}{link.}\footnote{\url{https://doi.org/10.5281/zenodo.10105278}} All interviewees consented to sharing an anonymized version of the interviews.

%% file: findings.tex
\section{Findings}\label{sec:findings}

Recall that \textbf{RQ1} focuses on intentionally designed and formalized accountability mechanisms, and \textbf{RQ2} seeks to identify organic or ``home-grown'' accountability practices outside the organizational influence. The latter naturally arises from the team's social dynamics and internal influences.

\begin{figure*}[bht]

\includegraphics*[trim=1cm 1.5cm 1cm 1cm, clip, width=.75\textwidth]{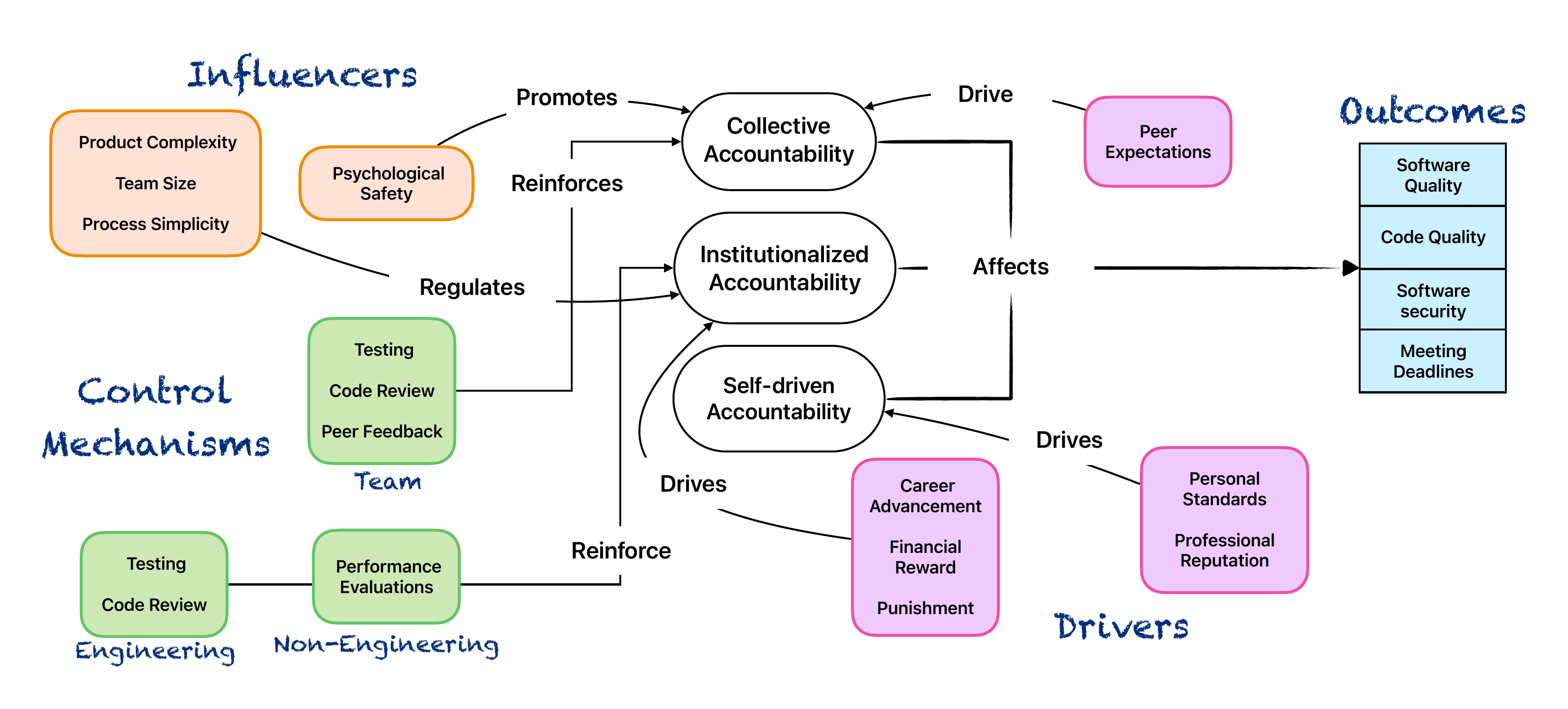}

\caption{Conceptual model for key building blocks in software engineering accountability. Pink nodes indicate Drivers of accountability; green nodes indicate Control Mechanisms to enforce accountability; Orange nodes indicate contextual Influences, which regulate accountability; blue nodes indicate Outcomes sought by the organization and the team by cultivating a sense of accountability. We describe each node type in the text.}

\label{fig:conceptual}

\Description{Conceptual model of accountability. Boxes connected in ways described in the text}

\end{figure*}

Fig. \ref{fig:conceptual}) presents the conceptual model drawn from our second cycle Pattern Codes. The model presents relationships, grounded in the data, that exist between distinct conceptual entities. It shows Influencers, Drivers, Control Mechanisms, and Outcomes. Influencers (orange) promote (grassroots) or regulate (institutionalized) accountability. Drivers (pink) are strategies to cultivate accountability. Control Mechanisms (green) enable teams and organizations to effectively manage and direct accountability processes. Finally, various forms of accountability — self-driven, collective (team), and institutionalized — affect organizational and individual Outcomes (blue), such as code quality (whether the ``code I wrote'' or the code of the entire product) or meeting a deadline.

We confirmed two distinct forms of accountability within software engineering teams: \emph{institutionalized} accountability (i.e., \textbf{RQ1}) and \emph{grassroots} accountability (i.e., \textbf{RQ2}). We found that grassroots accountability emerges organically within the team. This grassroots accountability is driven by factors like peers' expectations, personal standards, and intrinsic motivation, fostering a sense of collective or self-driven accountability in the team. Grassroots accountability tends to reward and nurture intrinsic motives, such as maintaining self-image within the team and reciprocity with peers. Institutionalized accountability, by contrast, relies on formal mechanisms and processes to exercise and formalize accountability, such as performance reviews, and more often uses traditionally known reward and punishment techniques, such as financial incentives, career advancement, and denial of promotion.

\subsection{Outcomes}

Institutionalized accountability drives toward certain outcomes. Our interviewees reported being individually accountable to the organization for software quality, code quality, software security, and meeting the project's deadlines.

While software quality refers to meeting the expectations of correctness and business requirements, code quality is more concerned with the internal features of the code, like \emph{``readability''} (P4, P11, and P12), and code modularity (P5, and P12). This is in line with Alami and Krancher findings, which suggest that practitioners prefer to delineate between ``external'' and ``internal'' aspects of software quality \citep{alami2022scrum}. While ``external'' aspects are usually defined from an end user perspective, code quality is an ``internal'' aspect of quality defined from a software engineering point of view \citep{alami2022scrum}.

For example, when P11 was asked why he feels accountable for the quality of his code, he stated, \emph{``so if, let's say we have people of 20 developer in my team, okay. And if my code quality is good, my image of the developer is good. Even if there is any certain bad conditions ... you have to cut developers out of by 20\% in your team, okay, I feel that my I should be amongst the good ones that my number doesn't come in top bottom 20\% of the developer in the team. That's what I feel''} (P11). P11's statement suggests that he perceives code quality as closely tied to his professional image and, subsequently, job security. The punishment-based strategy, i.e., losing one's job if underperforming, drives him to consistently deliver high-quality code to maintain a positive professional image and remain employed. He further explains that \emph{``good quality''} alone is not sufficient; meeting the deadlines is also an expectation. \emph{``... you have to develop something if you are not on the timeline? And but you have written the code good quality, then it also be not okay. Because you have to give it the delivery to the delivery timely''} (P11). P10 echoed this behavior. When asked why he cares about his work, he replies: \emph{``only way to progress further in your career is by caring about your work, caring about your product as well. Because no manager will come to you for promotion''} (P10). P11 and P10 accounts show a behavior towards maintaining high standards in software and code quality, driven by job security and career advancement. This behavior exemplifies the impact of institutionalized accountability, effectively aligning engineers' motivation with the organization's expectations for quality.

Engineers feelings of accountability for the team's outcomes are also driven by intrinsic motives and peers' expectations. For example, P9 passionately declares his accountability to the quality of the product and the product itself. He states: \emph{``I actually feel accountable for the whole, you know, whole software, actually ... I feel like it's my responsibility to do my best in order to produce something with quality''} (P9). P12 explains that she feels personally accountable for her code quality because it is culturally driven by her team's expectations. She explains: \emph{``so talking about that if I am accountable to the code quality, because suppose some other developer takes over me, or maybe someone has to extend the functionality, it should be very quick. And like it should be in a modular fashion. So, that is why we focus on code quality in the platform team''} (P12). P1 expresses a sense of personal accountability for quality and security, \emph{``quality and security, accountability, probably everything that I'm doing that could that can be critical in this condition, especially in the security department ... if if something, for example, obscures that our secure security critical ..., then I think I would feel guilty''} (P1).

While for P9, accountability seems to be a manifestation of intrinsic motivation, P1 expresses a sense of personal accountability driven by strong personal standards. The feeling of guilt shown by P1 indicates that he associates accountability with his personal responsibility to meet security requirements. P12's account demonstrates a collective sense of accountability within the team, where each member recognizes the significance of the quality of their code in facilitating future development efforts by ensuring a well-structured, maintainable, and extensible codebase.

\subsection{Drivers}

\subsubsection*{Institutionalized accountability drivers.}

Organizations intentionally implement strategies to drive accountability among software engineers. These formalized strategies shape engineers' commitment to achieving the intended outcomes. While rewards reinforce positive performance, punishment is a deterrent, discouraging engineers from underperforming by making them aware of the potential negative consequences.

\noindent \textbf{\textit{Financial rewards.}} Some of our interviewees expressed a motivation to meet established expectations, driven by the prospect of financial rewards. When asked what rewards a team member would receive for being accountable for the outcomes of their work, P6 mentions financial incentives, specifically the annual bonus tied to performance reviews. He states: \emph{``... obviously the bonus as well ... the bonus for me is something that I'm driven towards performing''} (P6). In a similar vein, P11 explains that his motivation to write \emph{``good quality code''} is not only rooted in maintaining his professional image, but also in the potential for financial growth, such as pay raises or promotions. He explains: \emph{``if I'm working somewhere, I think that I should ... getting good results is writing good quality code, then only I can get it as a pay raise or promotion''} (P11). Financial rewards emerged as drivers to uphold software engineers' accountability to meet established expectations. Our data show that these formalized incentives have far-reaching implications, extending to engineering outcomes such as software quality---an aspect traditionally perceived purely from an engineering perspective.

\noindent \textbf{\textit{Career advancement.}} Organizations reward meeting expectations for the intended outcomes with career advancements. Some of our interview data shows that the drive to receive advancements in engineers' careers is closely tied to demonstrating a strong sense of accountability. For example, a P10 account indicates accountability for one's work, and the product is a requirement for career advancement. He explains: \emph{``... only way to progress further in your career is by caring about your work ... this is the way I've seen people doing it by caring about their work ... The people who are like progressing further in the in the career, they always care about their work and their product. They are never the ones who ignore problems, they are the first one to jump in and solve them''} (P10). P10's statement demonstrates the connection between accountability and career growth. It underscores the importance of proactively caring about one's work and product as a key driver for professional growth. Similarly, when P12 was asked why she cares about the consequences of underperforming, she replied: \emph{``I care about consequences because... then we'll get a good salary. And I, my family, would have food''} (P12).

\noindent \textbf{\textit{Punishment.}} Organizations employ punishment-based strategies for underperforming team members to incentivize them to meet or exceed the established expectations for intended outcomes. Punishment-based strategies appeared in several interviewees' accounts, e.g., P8, P10, and P11. They range from corrective measures, such as \emph{``performance improvement plan''} to loss of employment. When P11 was asked about the consequences of not meeting expectations (in the case of his team \emph{``code quality''}), he replied: \emph{``... if this person is not improving, they will put him into the performance improvement plan, okay. There are certain days another performance improvement plan that they teach you about writing good quality of code writing better code. And if you get past from that performance improvement plan, then you can again join the team, otherwise you would get fired''} (P11). P10's company has a similar strategy: \emph{``either they [under performing team members] change or they quit the company that's how it happens. There's no scenario where you don't change and stay in the company''} (P10). Punishment based strategies demonstrate that, in some instances, organizations are committed to driving accountability among software engineers by implementing clear consequences for failing to meet expectations.

\subsubsection*{Grassroots drivers}

\textit{\textbf{Peers' expectations.}}

Peers' expectations are the implicit or explicit standards, behaviors, and performance criteria that software engineers perceive their colleagues to hold them accountable for. When peers' expectations are shared, they create a bond of answerability amongst team members, fostering a sense of collective accountability. Expectations are sometimes shared through knowledge-sharing processes, either formal or informal. P7's statement bluntly supports these findings: \emph{``I follow good technical practices in order not to disappoint my colleague during the code review''} (P7). Several interviewees conveyed that documented standards and guidelines are not always the norms, but rather team-negotiated expectations are followed, e.g., \emph{``talking to each other’s what quality should be has more importance to us''} (P9). Once these expectations are shared, they create a sense of collective accountability where software engineers feel compelled to reciprocate with their peers. P11 explains: \emph{``I have to write good quality code, good code, the good code, which is working and my teammates are happy with as per our agreed expectations''} (P11).

Grassroots is also rooted in a sense of self-driven accountability. Our interviews' data show that self-set personal standards and professional reputation are the main intrinsic drivers of self-driven accountability. The desire to maintain the reputation of a \emph{``good''} engineer (e.g., P11, P6), and self-set standards drive software engineers to feel accountable for the outcomes.

\noindent \textbf{\textit{Personal standards.}} Software engineers align their personal standards with the expectations they set for their own work. In this quote, P5 highlights that their code quality is important to them: \emph{``.. the quality is pretty good because it's important to me''} (P5). Similarly, P8 explains that quality is his own expectation: \emph{``I want to do it, like the highest quality that whatever I can deliver ... I mean, personally ... I want to build like the best quality that possible ... beyond expectation''} (P8). These personal standards are invested in the intended outcomes, contributing to enhanced results.

\noindent \textbf{\textit{Professional reputation.}} Software engineers also like to maintain their professional reputation within the team and the organization, and they associate this image with the quality of their work. Meeting established expectations is an investment in one's professional reputation. This drive has emerged as an intrinsic quality because our interviewees associated it with their \emph{``integrity''} (e.g., P7 and P11), \emph{``pride''} (e.g., P3 and P7) in their work, and being perceived as \emph{``good''} engineers (e.g., P8 and P12). P6 explains a deliberate behavior to maintain his image and to be known for the good things: \emph{``you want to be known for good things, not for the bad things ... you caught some bug and then you want to be known by the VP. You don't want to be known by the VP for your product broke''} (P6). P7 states: \emph{``Yeah. Your integrity will give you this professional image that will you preserve for the rest of your life, I guess''} (P7). These findings attest that this internal compass of integrity and pride in one's work transcends external expectations. The professional ethos of software engineers influence engineering's outcomes such as code quality and software security through a sense of accountability.

Our findings discern two pivotal constructs: accountability, a sense of obligation or readiness to justify actions to a specific entity or group; and motivation, referred to here as drivers (either institutionalized or innate), which are internal forces that propels individuals towards certain actions or behaviors, including their sense of accountability towards their teams' and individual's outcomes. While interconnected, our findings elucidate distinct roles each plays within software engineering teams. While the drivers are antecedents to the this sense of accountability, the later assumes a mediating role to influence the outcomes (e.g., code quality, and meeting deadlines).

\subsection{Control Mechanisms}

Control mechanisms are the various processes, tools, and procedures put in place by the organization or embedded within the software engineering practices, to ensure that individuals are accountable for meeting their responsibilities and adhering to established expectations. 

\subsubsection*{Institutional control}

Our interviewees reported non-engineering and engineering control mechanisms used in their organizations to control institutionalized accountability. For non-engineering mechanisms, performance evaluation was cited more frequently by our interviewees (e.g., P2, P6, P8, and P10). For engineering mechanisms, testing and code review are mostly used.

\noindent \textbf{\textit{Performance evaluation.}} Our interviewees reported that they undergo periodic evaluations where their performance is assessed against set expectations. Some metrics used in these assessments include the number of defects reported against the software engineer's code and feedback from business stakeholders and peers. P10 explains how this process works in his organization: \emph{``... there's a different process of accountability is that every semester after our first semester ends, there is a review of employee ... So this is the review with your manager. And your teammates also comment on your performance here ... So these people comment on my work that okay ... And then my manager will get this and will show anonymous report to me saying that this is the final thing. In this process, the manager will also go in and look at the code and the designs ... to take a final judgment, even if someone is saying this was good''} (P10). These periodic assessments drive accountability because they demand answerability from the organization for performance against predefined expectations.

\noindent \textbf{\textit{Testing.}} We use ``testing'' as an umbrella term to refer to various software testing techniques, such as end-user testing and continuous integration processes. This allows us to treat these processes as a single concept due to their shared primary intent, which is the early identification of defects. When P7 was asked how his accountability (i.e., \emph{``corrctness''} of functionality) is controlled, he replied: \emph{``I do a lot of testing personally. And then I have my colleague review the code''} (P7).

\noindent \textbf{\textit{Code review.}} Code review was extensively discussed by our interviewees as a control mechanism to reinforce accountability. Every interviewee (except P6) reported code review being practiced in their teams. Code review, as talked about in our data, seems to be a powerful accountability check. For example, P2 equates it to peers' \emph{``judgement''}. He states: \emph{``I mean, this [code review] is like, I mean, that's also like judgement''} (P2). When P11 was asked whether code review has a strong impact on making team members accountable for their work, he answered: \emph{``Yes, code review is a big thing ... if there is any bugs or any mistakes, I am also accountable for those mistakes''} (P11). Code review emerged in our data as an important mechanism for reinforcing accountability in software engineering teams, holding team members accountable for meeting quality expectations.

\subsubsection*{Grassroots control mechanisms}

The green box labeled ``Team'', in our model (Fig. \ref{fig:conceptual}), aggregates the mechanisms used at the team-level to control the collective sense of accountability. In addition to testing and code review, peers' feedback, whether part of code review processes or formal or unsolicited, drives software engineers to meet their peers' expectations, which in turn reinforces their sense of collective accountability.

\noindent \textbf{\textit{Peers' feedback.}} Our interviews' data indicate that when peer feedback is integrated into the development process, it fosters a sense of shared accountability within the team. Software engineers expect that their code will be reviewed, motivating them to meet their peers' expectations. P9 echoes this claim: \emph{``we are always all of us are you know, getting some feedback on our code. So even our team leader ... it's just a way to improve the quality''} (P9). P12 explains that she thrives to meeting her mentor's quality expectations, and she finds the positive visual clues communicated by her mentor emotionally rewarding. She said: \emph{``it motivates me to create the better code because in reviews, I get praise, for sure ... when we had a code review, I saw on his face like he was very happy because it was my little more effort that counted''} (P12). This positive reinforcement through feedback creates a desire amongst software engineers to repeatedly receive this reward (peers' appraisals). P11 explains a \emph{``good feeling''} when his peers appraise his code quality: \emph{``[it feels] so good, if any other developer also praises me that I have written a good code''} (P11). This positive reinforcement through feedback promotes a desire among software engineers to consistently seek recognition from their peers for the quality of their work, reinforcing their accountability to meet expectations.

\subsection{Influencers}

\subsubsection*{Institutional}

We identified three key influencers that shape institutionalized accountability: product complexity, product and process simplicity, and team size.

\noindent \textit{Product complexity.} For example, P10 explained that his team develops a complex product. P10 explains that it takes three to four days to run automated regression tests, \emph{``so it takes three days to run all the tests, three to four days''} (P10). P10's organization prefers punishment-based strategies. The need to implement institutionalize accountability is stronger when the product developed by the team is complex. P10 describes his company's stance if engineers appear indifferent toward the product: \emph{``if they [software engineers] don't care, the manager will scold them and make them care''} (P10).

\noindent \textit{Simplicity.} Teams with simple and predictable outcomes and processes seem to have a lesser need for formal accountability. For example, P4 describes his team development process as \emph{``simple''}; he elaborates: \emph{``focus on database extraction... the software requires minimal customization''} (P4). He then describes the deliverable: \emph{``it is quite simple in the output; the outcome is quite obvious''} (P4). This simplicity renders formal accountability unnecessary: \emph{``there is not much assessment of our work''} (P4).

\noindent \textit{Team size.} Small teams have a lesser need for rigorous and formal accountability processes. When P4 was asked whether his team of four engineers has formal rules and guidelines to assess meeting expectations, he replied: \emph{``we are more informal ... because we are a small team ... we tend to chat and we handle the things ... without going my emails, maybe because we are close in an open space''} (P4). He further explains that their manager replies with trust instead of control: \emph{``basically [our manager] trust us even he does not have a full view of the technical issue''} (P4).

In synthesizing these insights, we note that institutionalized accountability in SE is significantly influenced by contextual factors, as highlighted by the experiences of P10 and P4. Complex products necessitate stringent, often punishment-based accountability measures. Conversely, simpler projects and smaller team sizes tend to rely less on formal accountability structures and more on trust and informal communication. This variance implies that accountability mechanisms are adaptable within different SE contexts.

\subsubsection*{Psychological safety (PS)}

PS refers to the collective perception held by individuals that the workplace environment is conducive to interpersonal risk-taking without fear of negative consequences \citep{edmondson1999psychological}. PS was reported by our interviewees as a collective behavior within their teams, promoting their sense of shared accountability. They described this sense of safety in their teams when the absence of blame for mistakes is exercised and instead constructive approaches to mitigating errors and learning from them take place. P3 explains: \emph{``we don't try to punish anybody. We don't try to finger point and we don't try to blame. What we do is we sit down and we all we all get involved and asks, Where do we go wrong? What what happened? Why did this happen? How do we prevent it from happening again?''} (P3). This feeling of safety is conducive to a sense of togetherness, which is subsequently translated to a sense of collective accountability towards the outcomes. When P9 was asked why his team does not blame for mistakes, he answers: \emph{``we are just all looking to help each other to understand each other better, and thus produce good software ... we had a standard ... and somebody's not following this right away, you're just give him a friendly reminder ... you don't have to worry or be scared to make mistakes overall''} (P9). P9's team hold themselves and each other accountable for upholding these agreed-upon standards, yet in a psychologically safe approach.

%% file: discussion.tex
\section{Discussion}\label{sec:discussion}

In this section, we discuss the alignments and delineations from the existing theoretical framework, the implications for practice, and suggestions for future work.

\paragraph{\textbf{Alignments.}} Some of our findings resonate with the theoretical framework presented in Section \ref{sec:theory}. In SE environments, accountability is manifested through processes, mechanisms, and practices well supported theoretically. Software engineers periodically engage in an account-giving process to evaluate their performance against established expectations set by the organization or their teams. This account results either in rewards or sanctions \citep{hall2003accountability,frink1998toward,frink2008meso}. This may imply that SE environments adhere to the typical behavior of social systems. Similarly to other social systems, SE environments are akin to leveraging accountability mechanisms to control and mitigate outcomes' failures.

Our findings show that human and psychological qualities like feeling accountable have far-reaching implications for engineering outcomes such as code quality and software security. As humans, software engineers are motivated to see themselves in a positive light among their peers and the organization. This motivation is channeled into a feeling of accountability towards the qualities of the outcomes.

\paragraph{\textbf{Delineations.}} The internalization of accountability, which in turn is manifested in the quality of engineering outcomes, is not only catalyzed by institutionalized practices but also by the engineers' intrinsic motivation and a positive professional identity among their peers. While the existing theoretical framework posits that accountability drives are either formal \citep{frink1998toward,frink2008meso} or informal \citep{zellars2011accountability,romzek2012preliminary,romzek2014informal}, our study reveals that software engineers innate qualities, mainly their own set standards and professional reputation, also nurture their feeling of accountability.

Informal accountability is well-grounded in existing theories \citep{zellars2011accountability,romzek2012preliminary,romzek2014informal} and has been attributed to the shared norms of the group \citep{romzek2012preliminary}. However, it assumes that the rewards are similar to formalized accountability. In SE environments, software engineers feel accountable towards their peers' expectations to receive appraisals from their peers. In addition, sanctions at the team level are avoided. Instead, our work shows that the SE team prefers a psychologically safe approach to underperforming.

The existing theoretical framework also assumes that accountability is controlled predominately by performance evaluation and other informal methods \citep{gelfand2004culture,frink1998toward,romzek2014informal}, which is not always the case in SE. Our findings show that accountability arises not only from these formal and informal evaluations but also from the very nature of the engineering practices employed. Testing and code reviews, for instance, are integral to the SE process, and they inherently promote a sense of accountability towards the outputs they evaluate. They do so not solely through oversight but by fostering an environment where scrutiny and peer assessment are normalized, expected, and even welcomed.

The intrinsic qualities and how they influence the feeling of accountability towards the product unveil a craftsman quality amongst software engineers. This has been corroborated by studies in technical excellence in agile teams \citep{alami2021agile,alami2022journey}. Alami et al. reported that engineers in agile environment adopt a craftsman-like approach to achieving technical excellence, e.g., better software quality, and sustainable software design \citep{alami2022journey}. This craftsman-like mindset to software development suggests a model of accountability beyond adhering to the expectations; it is also about a personal commitment to the excellence of the software. 

\paragraph{\textbf{Implications on practice.}}

The findings that software engineers’ accountability is influenced not only by institutionalized mechanisms but also by peers' expectations and intrinsic motivation necessitate a re-examination of motivation strategies within SE organizations. While traditional accountability control mechanisms like performance evaluation will remain, our study shows that they can co-exist with peer-driven and intrinsic motivation-driven accountabilities. Our findings show that software engineers have a natural inclination towards peer-driven and intrinsic motivation-driven accountabilities. Organizations should capitalize on these qualities.

%% file: validity.tex
\section{Trustworthiness}\label{sec:validity}

In addition to \textit{triangulation}, \textit{saturation}, {\textit{peer debriefing}, and providing a \textit{thick description} (reported in Sect. \ref{sec:methods}), we carried out a \textit{member checking} activity to ensure our study's trustworthiness \citep{miles2014qualitative}.

\textit{Member checking}: We sought feedback from our interviewees on our findings. We documented our Pattern Codes in a separate document for each interviewee. We shared the link with each interviewee in Prolific and asked them to comment. Ten interviewees provided feedback. We paid an additional \textsterling5 for this effort. No major objections to our interpretations were reported. Some interviewees sought further clarifications prior to providing their support for the findings.

\textit{Triangulation}: We utilized data and Frink and Klimoski \cite{frink1998toward} theoretical framework to validate our findings. Upon the completion of our inductive analysis, we mapped our findings to the existing theoretical framework (Sect. \ref{sec:theory}) to identify support for similarities and discrepancies in our findings (see Sect. \ref{sec:discussion}).

\textit{Peer debriefing}: During the analysis process, the results were consistently discussed and scrutinized by two authors. The participation of two authors in the coding process helped minimize researcher biases. \citep{miles2014qualitative}.

%% file: conclusion.tex
\section{Conclusion}\label{sec:conclusion}

Our study sheds some light into the sociocultural underpinnings of accountability in the context of SE. In SE, the accountability of engineers extends beyond formalized processes and structures into the social and informal constructs, like meeting peers' expectations and the engineers' own innate motives. Our findings also show that a social and human quality such as feeling accountable can have far reaching implications on outcomes always considered purely engineering, such as code quality and software security. Our work contribute to the discourse that the development of software engineering as a discipline should not only be technically proficient but also socially aware.